\documentclass[11pt,a4]{article}

\usepackage{graphicx,floatflt,amssymb} 
\usepackage{epsfig} 
\usepackage{graphics} 
\usepackage{psfrag}
\usepackage[dvips]{feynmf}
\usepackage{sparticles}
\input epsf
\def\beq{\begin{equation}}
\def\eeq{\end{equation}}
\def\bea{\begin{eqnarray}}
\def\eea{\end{eqnarray}}
\def\bq{\begin{quote}}
\def\eq{\end{quote}}
\def \lsim{\mathrel{\vcenter
     {\hbox{$<$}\nointerlineskip\hbox{$\sim$}}}}
\def \gsim{\mathrel{\vcenter
     {\hbox{$>$}\nointerlineskip\hbox{$\sim$}}}}
\def\gappeq{\mathrel{\rlap {\raise.5ex\hbox{$>$}}
{\lower.5ex\hbox{$\sim$}}}}
\def\lappeq{\mathrel{\rlap{\raise.5ex\hbox{$<$}}
{\lower.5ex\hbox{$\sim$}}}}

\def\bea{\begin{eqnarray}}   
\def\eea{\end{eqnarray}}

\begin{document}
\vspace*{-1in}
\renewcommand{\thefootnote}{\fnsymbol{footnote}}
\begin{flushright}
\texttt{CI-UAN/04-16FT}\\
\texttt{LPT-Orsay/04-31}\\
\texttt{CUPub/04-10}\\
\end{flushright}
\vskip 5pt

{\begin{center}
{\Large \bf A model for leptogenesis at the TeV scale}
\vskip 25pt
{\bf Asmaa Abada,$^{1,}$}\footnote{E-mail address:
abada@th.u-psud.fr} {\bf Habib Aissaoui$^{1,2,}$}\footnote{E-mail address:
aissaoui@th.u-psud.fr}
 {\bf Marta Losada,$^{3,}$ }\footnote{E-mail address:
malosada@uan.edu.co} 
 \vskip 10pt  
$^1${\it Laboratoire de Physique Th\'eorique, 
Universit\'e de Paris XI, \\B\^atiment 210, 91405 Orsay Cedex,
France} \\
$^2${\it Laboratoire de Physique Math\'ematique et Physique Subatomique\\
Mentouri University, Constantine 25000 Algeria }\\
$^3${\it Centro de Investigaciones, 
Universidad Antonio Nari\~{n}o, \\ Cll. 58A No. 37-94, Santa Fe de Bogot\'{a},
Colombia}
\end{center}}
\vskip 20pt
\vskip 20pt
{\centering {{\bf Abstract}}\\}
We consider the mechanism of thermal leptogenesis at the TeV scale in
 the context of
an extension of the Standard Model with 4 generations and the inclusion
of four right-handed 
Majorana neutrinos. We obtain  a value for the baryon asymmetry of the
 Universe in accordance with
observations by solving the full set of coupled Boltzmann equations for this
 model.

\begin{quotation} {\noindent 
\vskip 10pt 
\noindent PACS number(s):~12.60.Jv, 14.60.Pq, 11.30.Fs\\
}

\end{quotation} \noindent{Leptogenesis, Neutrino Physics, Baryon Asymmetry of
the Universe.}

\vskip 20pt  

\setcounter{footnote}{0} \renewcommand{\thefootnote}{\arabic{footnote}}


 \newpage

\section{Introduction}

The attractive mechanism of thermal leptogenesis has been thoroughly investigated in
recent years. Most of the focus has been on the extended model of the Standard Model (SM)
with 3 right-handed (RH) neutrinos \cite{FY}. This simple model which can explain the baryon
asymmetry of the Universe (BAU) and provide appropriate  values for neutrino masses and 
mixing angles 
 is based on the out-of-equilibrium decay of
a RH neutrino, in a CP violating way, such that an asymmetry in the leptonic decay products
is produced. The leptonic asymmetry is then converted into a baryonic asymmetry
due to $(B+L)$-violating sphaleron interactions which are in equilibrium above
 the electroweak breaking scale.

From the experimental point of view the recent best fit values
of the ratio of the baryon  to  photon density $\eta =
\frac{n_B}{n_{\gamma}}$,
\bea \eta =
6.5^{+0.4}_{-0.3}\times 10^{-10}. \eea
is given by WMAP \cite{WMAP}.

The typical scale at which the standard scenario described above occurs 
corresponds to
 $10^{10}-10^{15}$ GeV, and the induced neutrino masses arise
 via
the see-saw mechanism  \cite{Ramond, yanagida2, MOHAPATRA-SENJANOVIC}  
with a common interaction being
at the origin of the asymmetry  and mass generation.

Recently, detailed calculations have been performed solving the corresponding 
Boltzmann equations of the 
system taking into account the different interactions which contribute both to 
the production and potential washout of the leptonic asymmetry.  
The implications of these analyses have provided strong constraints on
the values of the light left-handed neutrino masses  $\sum_i m_i \sim
0.2$ eV \cite{bdp1,bdp2,bdp3} and a lower bound on the mass of the 
lightest RH neutrino
  $M \gsim 10^8 $GeV \cite{bdp1,di,branco}. The more recent work of ref.
   \cite{pilaftsis} has included the
effect of scattering with gauge bosons which had been neglected 
before and in ref. \cite{Giudice} a
comprehensive calculation including effects from renormalization group (RG)
running, finite temperature background effects, other possible decays producing
a  leptonic asymmetry (decay of a scalar particle), a correction to the 
appropriate subtraction
that must be performed for scattering via N-exchange was presented.
A thorough recalculation of this last issue was done in ref. \cite{bdp04}, 
coinciding with
ref. \cite{Giudice} that the washout contribution from this type of 
diagram had been overestimated.

The less standard scenario of leptogenesis occurring at the TeV scale
has been investigated for example in references 
\cite{al,moriond,Senami,Senami2,Boubekeur,hambye2,march-russell,pilaftsis2}.
In particular, in ref. \cite{al} a TeV scale model which is a simple extension
of the Fukugita-Yanagida (FY) model including one more generation to the SM and four gauge singlets was discussed and estimates of the produced
CP-asymmetry and final baryon asymmetry were obtained.
Both supersymmetric and non-supersymmetric versions of the model were analyzed.

In this paper we focus on this model of ref. \cite{al} and solve the 
corresponding Boltzmann equations (BE)
in order to reliably obtain the final values of the asymmetry and determine
the region of parameter space which is available for the model.

We dedicate section 2 to a careful discussion of the way we set up our 
calculations of the Boltzmann equations
making emphasis on  a few subtle points. 
In section 3, we recall the main features of the model and establish our 
notation.
Section 4 presents the Boltzmann equations valid for this model.  
Section 5 is devoted to our  results and 
 conclusions.

\section{Preliminary Considerations}

In reference \cite{Barbieri} it was pointed out that the relevant object that
should be studied in thermal leptogenesis is the density matrix $\rho$ 
\cite{sigl}. In the FY model $\rho$ is a
$(3\times 3)$ matrix
with all entries different from zero for temperatures greater than $10^{12}$GeV.
However, as the temperature decreases and each of the  different interactions
 defined by a specific
 charged Yukawa coupling
enters equilibrium  the corresponding  off-diagonal elements 
of the density matrix go to zero.
In our model at the TeV scale all of the charged Yukawa couplings are 
in equilibrium 
and thus $\rho$ will 
be a diagonal matrix. The matrix $\rho$ is normalized such that,
\bea
Tr \rho = \sum_{\alpha} Y_{L_{\alpha}} 
\equiv \sum_{\alpha} \left(Y_{\ell_{\alpha}}
-Y_{\bar{\ell}_{\alpha}}\right)\ .
\eea
where the sum over $\alpha$ indicates a sum over flavour.

Another important issue is the relationship between the produced lepton asymmetry and
the induced $B-L$ asymmetry which is conserved by sphaleron interactions. Again a careful consideration
of the temperatures and which interactions are in equilibrium changes the equation relating these
asymmetries, see reference \cite{Barbieri}.
In general the procedure that is employed is through the establishment of
equations for the chemical potentials reflecting which interactions are in equilibrium. Thus the chemical equilibrium equations allow us to write the asymmetries in number densities in terms of some specific asymmetry, say $Y_{L_{\alpha}}$.
For the Fukugita-Yanagida model then generically there would be a $(3\times 3)$
 matrix $A_{\alpha\beta}$ of the form
\cite{Barbieri}
\bea
Y_{B/3-L_{\alpha}} = A_{\alpha\beta} Y_{L_{\beta}},
\label{generic}
\eea 
and the numerical values for the elements of $A$ are determined from the equations for the chemical potentials.
If all interactions related to charged Yukawa couplings are in equilibrium, as well as sphaleron interactions,
the equation for the case of N generations simplifies to,
\bea
\frac{Y_B}{N} - Y_{L_{\alpha}} = 
\left(-\frac{4}{3N} + \frac{8}{3}\frac{1}{4N+2} \right) \sum_{\beta} Y_{L_{\beta}} - 3 Y_{L_{\alpha}},
\label{bli}
\eea
or adding over all generations
\bea
Y_{B-L} = -\left( {22N+13\over 6N+3} \right )\sum_{\alpha} Y_{L_{\alpha}}.
\label{bl}
\eea

Some comments are in order. In the standard scenario of the  FY model with three 
RH neutrinos,
where the production of the asymmetry occurs at some high scale for 
which not all charged Yukawa couplings are in equilibrium, one 
should use eq. (\ref{generic}) to relate the asymmetries. The point is that usually what is
solved in the system of coupled Boltzmann equations, under certain assumptions, 
is an equation for $Y_L =\sum_{\alpha} Y_{L_{\alpha}}$, see below for a discussion 
on this
 point, and then the asymmetry in $B-L$ is calculated using 
 eq. (\ref{bl}). From our previous discussion we see that it is not 
 quite correct to do this unless some approximations are done concerning the
 different $Y_{L_{\alpha}}$. 

For our model at the TeV scale
 \footnote{In this model, we denote by $\sigma$ the flavour of the fourth 
 leptonic generation.}, all charged Yukawa couplings are in equilibrium and 
 it would be  correct to
use eqns. (\ref{bli}) and (\ref{bl}). Another feature of our model is
that the Yukawa coupling of the heavy LH neutrino is of the order of $ 1$ to ensure that the masses of the fourth generation
leptons are heavy enough, and also to give an enhancement to the value of the 
CP-asymmetry produced in the model.
 However, then the 
scattering interaction which is associated to the  Yukawa 
coupling of the heavy LH neutrino will also be in 
equilibrium in our model. This 
changes the chemical equilibrium equations such that the relationship 
between  $B-L$ ($B$) and $L$ is modified to be:
\bea
Y_{B-L} =-\left({13\over 3} + {8\over 3}{(13-3N)\over(12N-2)}\right)
\sum_{\alpha=e,\mu,\tau} Y_{L_\alpha}=
 - \left({22N+13\over 6N-1}\right)\sum_{\alpha=e,\mu,\tau} Y_{L_\alpha}.
\label{finalbl}
\eea

Thus, as was estimated in reference \cite{al} there is no net asymmetry produced 
in the heavy left-handed lepton.
Consequently, we can put $Y_{L_{\sigma}}$ to zero and make the usual 
assumptions about the other  $Y_{L_\alpha}$
when constructing  the Boltzmann equation for $Y_L$ provided we use
 eq. (\ref{finalbl})
to relate the produced leptonic asymmetry to the final net $B-L$ 
asymmetry. Below, and from now on,  $Y_L$ denotes the sum over the asymmetries of only the three 
light leptonic flavours.

In reference \cite{Giudice} the effect of renormalization group (RG) running 
for masses and couplings was
included. In particular, one of the important effects is that the diagrams
 with $\Delta L =1$  proportional to gauge couplings
can become sizeable compared to those proportional to the 
top Yukawa coupling when considering values
for these couplings at the high scale at which the lepton asymmetry 
is being produced. In our model
at the TeV scale, we have checked that  this is a very small effect and
 we will not include RG effects in our analysis.
We also have not included the $\Delta L=1$  proportional
 to gauge couplings as we
are not in the resonant leptogenesis case of ref. \cite{pilaftsis}.

Also in ref. \cite{Giudice} thermal background effects have been included 
into the calculation, 
it was pointed out that numerically these effects are particularly relevant  
when correcting the propagator of the Higgs scalar field. We include only this thermal correction
into our calculation.

Another important point that should be mentioned, which is crucial for our model, is the fact
that the actual decay temperature at which the lepton asymmetry is produced is just above the
electroweak scale; thus one must be careful and check that for our choice of parameters the
 sphaleron interactions are still in equilibrium. Although, in the SM there is no first order
phase transition, just a crossover, and this is not modified for our model with four generations, 
sphaleron interactions are switched off for temperatures below $\sim 100$ GeV \cite{moore}.
A careful calculation of the chemical equilibrium equations in the region close to the 
electroweak phase transition was performed in \cite{mikkomisha}.

\section{The Model}

The relevant part of the  Lagrangian of the model we consider is given by, 
\bea L = L_{SM} + \bar{\psi}_{R_{i}} i \partial \! \! \! /~\psi_{R_{i}} -
\frac{M_{N_{i}}}{2}(\bar{\psi}_{R_{i}}^{c} \psi_{R_{i}} + h.c.) - 
(\lambda^{\nu}_{i\alpha} \bar{L}_{\alpha} \psi_{R_{i}} \phi + h.c.) \
,\label{lag} \eea where 
$\psi_{R_{i}}$ are two-component spinors describing the right-handed neutrinos 
and we define a  Majorana 4-component spinor, $N_{i} =  \psi_{R_{i}} + 
\psi_{R_{i}}^{c}$. Our index i runs from 1 to 4, and $\alpha=e,\mu,\tau,\sigma$. 
The $\sigma$ component of $L_{\alpha}$
corresponds to a left-handed lepton doublet which must satisfy the LEP
constraints from the Z- width on a fourth left-handed neutrino~\cite{pdg}. The 
$\lambda^{\nu}_{i\alpha}$ are Yukawa couplings and  the field $\phi$ is the
 SM Higgs boson
doublet whose vacuum expectation value is denoted by $v$.

We work in the basis in which the mass matrix for the right-handed neutrinos
$M$  is diagonal and real,
\bea M = diag(M_1,M_2,M_3,M_4) \eea and define $m_D = \lambda^{\nu} v$.

To leading order~\footnote{Given the
fact that in our model,  at least one of the  RH neutrino masses take values 
$\sim$ TeV,
 one has to go to
next order in the general  see-saw formula. However, these corrections are small and not relevant for our purpose.}, 
the induced see-saw
neutrino mass matrix for the  left-handed neutrinos is given by, 
\bea m_{\nu}  = \lambda^{\nu \dagger} M^{-1} \lambda^{\nu} v^2  \ .
\eea 
which is diagonalized by the well known PMNS  matrix.

We consider the
out-of-equilibrium decay of the  lightest of the gauge singlets $N_j$, which we
take to be  $N_1$.  The decay rate at tree-level 
 is given by
\bea  \Gamma_{N_j}=\frac{\left(\lambda^\dagger\lambda\right)_{jj}}{8\pi}{M_j}=
\sum_{\alpha=1}^{4}\frac{\left(\lambda^*_{\alpha j}\lambda_{\alpha j}\right)}{8\pi}{M_j} = 
{\tilde {m}_{j}^{(4G)}M_1M_j\over 8\pi v^2},\label{gammaN} 
\eea
where $\lambda \equiv \lambda^\nu$ and we have defined
\footnote{To simplify our notation, we write the sums over 
$\alpha=1$ to 4 denoting
the sums over all flavours $e,\mu,\tau,\sigma$, and 
sums over $\alpha=1$ to 3 to denote sums over the three lightest flavours. For
instance, $\tilde{m}^{(3G)}_j=\sum_{\alpha=1}^{3}\frac{\left(\lambda^*_{\alpha j}
\lambda_{\alpha j}\right)}{M_1}v^2$ is the usual effective mass parameter 
$\tilde
m_1$ defined in the literature for $j=1$.}
\bea \tilde{m}^{(4G)}_j=\sum_{\alpha=1}^{4}\frac{\left(\lambda^*_{\alpha j}
\lambda_{\alpha j}\right)}{M_1}v^2.\eea
To ensure an out-of-equilibrium decay of $N_1$  it is necessary that
$\frac{\Gamma_{N_{1}}}{H(T=M_{1})} \ll 1$, where H is the Hubble  expansion
rate at $T=M_{1}$. 
This
condition will impose an upper bound on the usual effective mass parameter 
$\tilde{m}_1$ which is defined in for example \cite{Plumacher}.

 The CP asymmetry calculated from the
interference of the tree diagrams  with the one-loop
diagrams (self-energy and vertex corrections)  is
\cite{roulet}
\bea \epsilon_i = \frac{1}{(8\pi)}\frac{1}{  
[\lambda^{\dagger}_{\nu} \lambda_{\nu}]_{ii}}
\sum_j {\mathrm Im}{ [\lambda^{\nu \dagger}
\lambda_{\nu}]^2_{ij}}\left[f\left(\frac{M_{j}^2}{M_{i}^{2}}\right) + g
\left(\frac{M_{j}^2}{M_{i}^{2}}\right)\right] \label{asym}\ , \eea 
where
\bea f(x) &=& \sqrt x[1 - (1+x)\ln\frac{1+x}{x}], \nonumber\\ g(x) &=&
\frac{\sqrt x}{1-x}. \eea 

In reference \cite{al} it was shown that
 the upper bound of the
CP-asymmetry produced in the decay of the lightest right-handed neutrino $N_1$
is  \bea |\epsilon_1| \lsim \frac{3}{8\pi} \frac{M_1 m_4}{v^2},
\label{cpbound} \eea where $m_4$ denotes the largest eigenvalue of the
left-handed neutrino mass matrix  $m_{\nu}$. Due to experimental constraints
 we
require that $m_4 > 45$~GeV, which implies that the bound on $\epsilon_1$ is
irrelevant. In general there will be regions of parameter space in which the
 produced CP
asymmetry can be very large, thus the final allowed region  of parameter space
could include areas in which the washout processes can be very large as well.
In reference \cite{al} possible textures for the  $\lambda^{\nu}$ 
matrix were given 
to illustrate
how to obtain masses consistent with low energy data and the appropriate 
amount of
CP asymmetry from the decay of $N_1$.

\section{Boltzmann Equations}

We will now write the Boltzmann equations for the RH neutrino abundance and
the lepton densities. With all the above arguments in consideration, 
the main processes we consider 
in the thermal bath of the early universe are decays, inverse decays 
  of the RH neutrinos, and the lepton number violation $\Delta L=1$  
  and $\Delta L=2$,
  Higgs and RH neutrinos exchange scattering processes, 
  respectively. 

In our analysis we stick to the case  where the asymmetry is due only
   to the decay of the lightest
    RH neutrino $N_1$ \footnote{We have checked that  the asymmetry produced by the
    decay of the second lightest RH neutrino is very tiny 
    in our model compared to the one produced by the decay of $N_1$.}.
      The first BE, which corresponds to the evolution of the abundance  
 of the lightest RH neutrino $Y_{N_1}$ involving  the decays,  
    inverse decays and $\Delta L=1$ processes is given by 
    \bea{dY_{N_{1}}\over dz}=-{z\over sH(M_1)}({Y_{N_{1}}
    \over {Y^{eq}_N}} -1)(\gamma_{_{D_1}}+\gamma_{_{S_1}}), \label{be1} \eea
     where $z=\frac{M_1}{T}$, $s$ is the entropy density and 
      $\gamma_{_{D_j}},\gamma_{_{S_j}}$ are the interaction 
     rates for the decay 
and $\Delta L=1$  scattering  contributions, respectively. \\ 
The BE for the lepton asymmetry is given by
 \bea {dY_L\over dz}=-{z\over sH(M_1)}
 \bigl[\epsilon_1 \gamma_{_{D_1}}( {Y_{N_{1}}\over Y^{eq}_N}-1)
 + \gamma_{_{W}} {Y_{L}\over Y^{eq}_L}\bigr], \label{be2} \eea 
 where $\epsilon_1$ is the CP violation parameter given by eq. (\ref{asym})
  and $\gamma_{_W}$, is a function of $\gamma_{_{D_j}}$ and $\gamma_{_{S_j}}$
   and 
  $\Delta L=2$ interaction rate processes, called the washout factor which is responsible 
  for the damping of the produced asymmetry. 
  In eqs. (\ref{be1}) and (\ref{be2}),  $Y^{eq}_{i}$ is the equilibrium number 
  density of
  a particle $i$, which has a mass $m_i$,  given by 
  \bea
  Y^{eq}_{i}(z)={45\over 4 \pi^4}{g_i\over g_*} \left({m_i\over M_1}\right)^2 
  z^2
  K_2\left({m_i z\over M_1}\right)\ ,
  \eea
  where $g_i$  is the internal degree of freedom of the particle 
  ($g_{_{N_i}}=2$, $g_{_{\ell}}=4$) and $g_*\simeq
 108$. 
   Explicit expressions of the interaction 
   rates $\gamma_{_{D_j}}$, $\gamma_{_{S_j}}$ and $\gamma_{_W}$
    are given below.\\
The reaction density $\gamma_{_{D_j}}$ is related to the tree level total decay 
rate  of the  RH neutrino $N_j$ of eq. (\ref{gammaN}) by
\bea  \gamma_{_{D_j}}=n^{eq}_{N_j}{\frac{K_1(z)}{K_2(z)}}\Gamma_{N_j},\eea
where  $K_n(z)$ are the $K$ type Bessel functions.\\
We define  the reaction density of any process $a+b\to c+d$ by 
\bea 
\gamma^{(i)}=
\frac{M_1^4}{64\pi^4}\frac{1}{z}\int_{\frac{(M_a+M_b)^2}{M_1^2}}^{\infty}
{dx~\hat{\sigma}^{(i)}(x)\sqrt{x}~K_1\left(\sqrt{x}z\right)},
\eea
where $\hat{\sigma}^{(j)}(x)$ are the reduced cross sections for the different 
processes which contribute
to the Boltzmann equations. 
For the $\Delta L=1$ Higgs boson exchange processes, we have
\bea 
\hat{\sigma}_{t_j}^{(1)}&=&3\alpha_{u}\sum_{\alpha=1}^{4}
\left(\lambda^*_{\alpha j}\lambda_{\alpha j}\right)
\left(\frac{x-a_{j}}{x}\right)^2 
=3\alpha_{u}\frac{\tilde{m}^{(4G)}_jM_1}{v^2}\left(\frac{x-a_{j}}{x}\right)^2,
\label{sigma1}\\
\hat{\sigma}_{t_j}^{(2)}&=&3\alpha_{u}\sum_{\alpha=1}^{4}
\left(\lambda^*_{\alpha j}\lambda_{\alpha j}\right)
\left(\frac{x-a_{j}}{x}\right)\left[\frac{x-2a_{j}+2a_{h}}{x-a_{j}+a_{h}}+
\frac{a_{j}-2a_{h}}{x-a_{j}}
\ln\left(\frac{x-a_{j}+a_{h}}{a_{h}}\right)\right] \nonumber\\
&=&3\alpha_{u}\frac{\tilde{m}^{(4G)}_jM_1}{v^2}
\left(\frac{x-a_{j}}{x}\right)\left[\frac{x-2a_{j}+2a_{h}}{x-a_{j}+a_{h}}+
\frac{a_{j}-2a_{h}}{x-a_{j}}
\ln\left(\frac{x-a_{j}+a_{h}}{a_{h}}\right)\right],\label{sigma2}
\eea
where 
\bea \alpha_u= \frac{Tr(\lambda_u^\dag\lambda_u)}{4\pi}\simeq \frac{m_t^2}{4\pi v^2}, 
\hspace{0.5cm}a_j= {\left({\frac{M_j}{M_1}}\right)^2}, \hspace{0.5cm}
a_h={\left({\frac{\mu}{M_1}}\right)^2},\eea
the infrared cutoff $\mu$ is chosen to be equal to $800$ GeV due to phenomenological
considerations. So we have  
\bea \gamma_{_{S_j}}=2\gamma_{_{t_j}}^{(1)}+4\gamma_{_{t_j}}^{(2)}.\eea
The expressions of reduced cross sections of $\Delta L= 2$ RH neutrino exchange processes, 
which  washout a part of
the asymmetry are given by
\bea \hat{\sigma}_{N}^{(1)}&=&\sum_{\alpha=1}^{4}\sum_{\beta=1}^{3}\sum_{j=1}^{4}
\left(\lambda^*_{\alpha j}\lambda_{\alpha j}\right)
\left(\lambda^*_{\beta j}\lambda_{\beta j}\right)
\it{A}_{jj}^{(1)}+
\sum_{\alpha=1}^{4}\sum_{\beta=1}^{3}\sum_{\stackrel{n<j}{j=1}}^{4}
Re\left(\lambda^*_{\alpha n}\lambda_{\alpha j}\right)
\left(\lambda^*_{\beta n}\lambda_{\beta j}\right) \it{B}_{nj}^{(1)} \label{sigmaN1} 
 \\
 \hat{\sigma}_{N}^{(2)}&=&\sum_{\alpha=1}^{4}\sum_{\beta=1}^{3}\sum_{j=1}^{4}
\left(\lambda^*_{\alpha j}\lambda_{\alpha j}\right)
\left(\lambda^*_{\beta j}\lambda_{\beta j}\right)
\it{A}_{jj}^{(2)}+
\sum_{\alpha=1}^{4}\sum_{\beta=1}^{3}\sum_{\stackrel{n<j}{j=1}}^{4}
Re\left(\lambda^*_{\alpha n}\lambda_{\alpha j}\right)
\left(\lambda^*_{\beta n}\lambda_{\beta j}\right) \it{B}_{nj}^{(2)} 
 \label{sigmaN2}
 \eea 
 where 
 \bea
 \it{A}_{jj}^{(1)}&=& \frac{1}{2\pi}\left[1+\frac{a_j}{D_j}+\frac{a_j x}{2D^2_{j}}-
 \frac{a_j}{x}\left(1+\frac{x+a_j}{D_j}\right)\ln\left(\frac{x+a_j}{a_j}\right)\right],
 \\
 \it{A}_{jj}^{(2)}&=&
 \frac{1}{2\pi}\left[\frac{x}{x+a_j}+\frac{a_j}{x+2a_j}\ln\left(\frac{x+a_j}{a_j}
 \right)\right],\\
 \it{B}_{nj}^{(1)}&=& \frac{\sqrt{a_n a_j}}{2\pi}
  \left[\frac{1}{D_j}+\frac{1}{D_n}+\frac{x}{D_jD_n}+
\left(1+\frac{a_j}{x}\right)\left(\frac{2}{a_n-a_j}-
\frac{1}{D_n}\right)\ln\left(\frac{x+a_j}{a_j}\right)\right.\nonumber\\
&+& \left.\left(1+\frac{a_n}{x}\right)\left(\frac{2}{a_j-a_n}-\frac{1}{D_j}\right)
\ln\left(\frac{x+a_n}{a_n}\right)\right], \\
 \it{B}_{nj}^{(2)}&=& \frac{\sqrt{a_na_j}}{2\pi}
 \left\lbrace\frac{1}{x+a_n+a_j}\ln\left[\frac{(x+a_j)(x+a_n)}{a_ja_n}\right]+
 \frac{2}{a_n-a_j}\ln\left(\frac{a_n(x+a_j)}{a_j(x+a_n)}\right)\right\rbrace,
 \eea
and
\bea
 D_j={\frac {\left( x-a_{j}\right)^{2}+a_{j}c_{j}}{x-a_{j}}},
\hspace{1cm} 
c_{j}=a_{j}\sum_{\alpha=1}^4,\sum_{\beta =1}^{3}
{\frac{\left(\lambda^*_{\alpha j}\lambda_{\alpha j}\lambda^*_{\beta j}
\lambda_{\beta j}\right)}
{64\pi^2}}, \eea
then we have 
\bea \label{gammaW}
 \gamma_{_W}=\sum_{j
=1}^{4}\left(\frac{1}{2}\gamma_{_{D_j}}+\frac{Y_{N_j}}{Y_{N_j}^{eq}}
\gamma_{_{t_j}}^{(1)}
+2\gamma_{_{t_j}}^{(2)}-{\gamma_{_{D_j}}\over 8}\right)+2\gamma_{N}^{(1)}+2\gamma_{N}^{(2)}
 \ .\eea

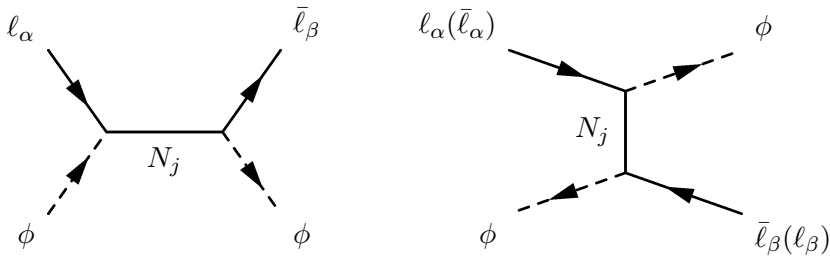
\begin{figure}[t] 
\begin{center}
\begin{tabular}{cccccccccccccccc}
\begin{fmffile}{one} 	  
  \fmfframe(1,7)(1,7){ 	
   \begin{fmfgraph*}(110,62) 
    \fmfleft{i1,i2}	
    \fmfright{o1,o2}    
    \fmflabel{$\phi$}{i1} 
    \fmflabel{$\ell_\alpha$}{i2} 
    \fmflabel{$\phi$}{o1} 
    \fmflabel{$\bar \ell_\beta$}{o2} 
    \fmf{scalar}{i1,v1}
    \fmf{scalar}{v2,o1} 
    \fmf{fermion}{i2,v1}
    \fmf{fermion}{v2,o2} 
    \fmf{vanilla,label=$N_j$}{v1,v2} 
   \end{fmfgraph*}
   }
\end{fmffile}
&&&&

\begin{fmffile}{two} 	  
  \fmfframe(1,7)(1,7){ 
   \begin{fmfgraph*}(110,62)
    \fmfleft{i1,i2}
    \fmfright{o1,o2}
    \fmflabel{$\phi$}{i1}
    \fmflabel{$\ell_\alpha(\bar\ell_\alpha)$}{i2}
    \fmflabel{$\bar\ell_\beta(\ell_\beta)$}{o1}
    \fmflabel{$\phi$}{o2}
    \fmf{scalar}{v1,i1}
    \fmf{scalar}{v2,o2}
    \fmf{fermion}{i2,v2}
    \fmf{fermion}{o1,v1}
    \fmf{vanilla,label=$N_j$}{v1,v2}
   \end{fmfgraph*}
   }
\end{fmffile}
\end{tabular}
\caption{ $s$ and $t$ channel $\Delta L=2$ Majorana exchange diagrams.} \label{feyn}\
\end{center}
\end{figure}

There are several important comments to make at this point.
 First of all,  notice that in eqs. (\ref{gammaN}), (\ref{sigma1}) 
and (\ref{sigma2}), we have summed over all states $\ell_\alpha$.
 However in eqs. (\ref{sigmaN1}) 
and (\ref{sigmaN2}) terms which correspond to 
$\beta=4$, (that is the $\sigma$ flavour, see figure \ref{feyn}),
 do not contribute to 
equation (\ref{be2}) because they are multiplied by $Y_{L_{\sigma}}=0$.
Indeed, as was mentioned before, the processes involving at the same time 
only this flavour in the initial and final state are in thermal equilibrium, 
and so there is no
asymmetry in this flavour. 
 Second,
as in the case of FY model, one can parametrize $\gamma_{_{D_j}}$ and
$\gamma_{_{S_j}}$ by $M_1$ and $\tilde{m}^{(4G)}_1$
 (see eqs. (\ref{gammaN}), (\ref{sigma1}), 
(\ref{sigma2})). 
Third, the $\Delta L=2$ processes can be divided and 
treated in two separate regimes. 
The contribution in the temperature range from $z=1$  to $z\simeq 10$
is proportional to $\sqrt{\tilde{m}^{(4G)}_1\tilde{m}^{(3G)}_1}$ 
instead of
$\tilde{m}_1$. 
For the range of
temperature $(z>>1)$, the dominant contribution at leading order 
would have been
proportional to 
 $\bar{m}^2 \equiv tr(m_{\nu}^{\dagger}m_{\nu})$, had the summation on
the index $\beta$ gone from 1 to 4.
 In addition, we emphasize that  the so-called RIS (real
intermediate states) in the $\Delta L=2$ interactions have 
to be carefully subtracted, to avoid
 double counting in the Boltzmann equations.
  In our calculation, we have followed 
ref. \cite{bdp04} where the
authors have shown how the appropriate ($-{1\over 8} \gamma_{_{D_j}}$ 
in eq. (\ref{gammaW}))
 subtraction must be done.
\section{Results and Conclusions}
\subsection{Constraints on the fourth generation}\label{4gen}
Our model is based on the addition of a fourth doublet of leptons and quarks 
(in order to be anomaly-free) to the particle content of 
the Standard Model, and  of four RH neutrino singlets. 
The masses and couplings of the singlets are constrained by the 
left handed neutrino masses (from solar, atmospheric and WMAP data) 
and the right amount of CP asymmetry
$\epsilon_1$.
Concerning the fourth generation of leptons and quarks,
 the direct constraint is that they must be heavier than $\frac{M_Z}{2}$. 
 There are
stronger constraints from electroweak precision tests: for example, 
the lower bound on the 
charged lepton $\sigma$ from LEP II  is approximately 80 GeV \cite{Maltoni}.
As is well known, the SM central value of the Higgs mass is lower than the direct 
lower
bound set by LEP II \cite{LEPII} and the existence of a fourth
generation is one possible way to increase the Higgs mass
\cite{Polonsky,OKUN1,Sultansoy}.
 The Z-lineshape versus fourth generation masses have been extensively 
 studied in
 for example ref. \cite{OKUN2} where we can see that the fourth left-handed
 neutrino is excluded at 95\% CL if its mass is lower than $46.7\pm 0.2$ GeV.
 In our analysis the left handed neutrino masses are always constrained  in
 order to fit the data (solar + atmospheric + WMAP) as well as this latter 
 bound for the 4th LH neutrino. 
 There are other constraints on the fourth generation, the
 generalized CKM matrix elements, etc.
  which can be found in \cite{pdg} and in ref. \cite{Frampton}, 
   where present and
  future experimental searches at Tevatron and LHC are widely discussed.
\subsection{Numerical solutions of BE}
In our numerical analysis we use one of the possible textures for the Yukawa matrix($\lambda_\nu$) in the four generation case which can produce the needed  amount of CP-asymmetry and fit  the neutrino data, \cite{al}: 
\bea \lambda_{\nu} = C \left(\begin{array}{cccc} 
\epsilon & \epsilon & \epsilon &\alpha\\ \epsilon & 1 & 1 & 0 \\ \epsilon & 1 & 1 & 0
 \\ \epsilon & 0 & 0 &1/C\\ \end{array}\right),\eea where $C, \epsilon $ and $\alpha$ are parameters ($\epsilon $ and $\alpha$ are complex numbers). 
This texture will induce to  first order the following mass matrix for the three  lightest LH neutrinos 
\bea m_{\nu} \propto \left(\begin{array}{ccc} \epsilon^2 &\epsilon & \epsilon \\ 
\epsilon & 1 & 1 \\ \epsilon & 1 &
1\end{array}\right). \eea
This is a typical texture for a hierarchical spectrum that has to fit the  oscillation data (atmospheric, solar, CHOOZ) and the constraints on the absolute mass. 

To show the feasibility of our scenario we chose different values of the parameters $C, \epsilon $, $\alpha$ and values of the RH neutrino masses, $M_i$, $i=1,4$ and we constrain the inputs given the data for the LH neutrino masses and the out-of- equilibrium condition. It is precisely a hierarchy in the Yukawa couplings which allows for an out-of-equlibrium decay without a suppression of the induced CP-asymmetry.

\begin{figure}[t]
 \begin{center}
 \includegraphics[width=12cm]{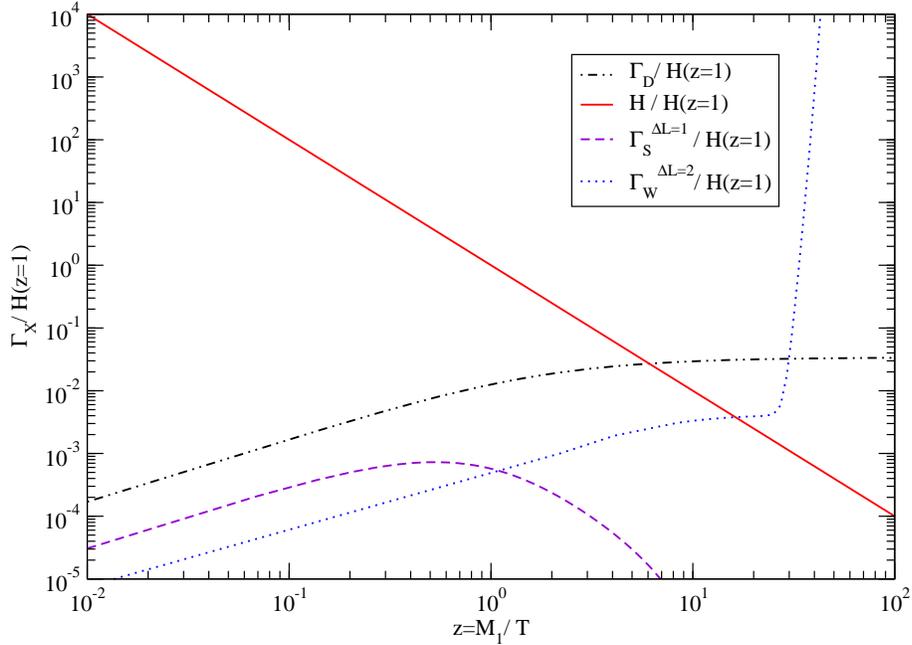}
\caption{Various thermally averaged reaction rates $\Gamma_X$ 
contributing to BE normalized to
the expansion rate of the Universe $H(z=1)$.} 
\label{rates4G}  
\end{center}

\end{figure}

In figure \ref{rates4G}, we illustrate for a given set of input parameters
  the different thermally averaged reaction rates contributing to BE 
  as a function
of  $z=\frac{M_1}{T}$: 
\bea
\Gamma_{_X}={\gamma_{_X}\over n^{eq}_{N_1}},\hspace{1.5cm} 
X=D,\ S, \ \Delta L=2 \, .
\eea It is clear from
  this plot that for this  set of parameters, and this is  true  for a wide 
  range of
  parameter space, all rates at $z=1$ fulfill the out-of-equilibrium 
  condition
  (i.e. $\Gamma_X<H(z=1)$), and so the expected washout effect due to the
  $\Delta L=2$ processes will be small.
The parameters chosen for this illustration are: $M_1=450$ GeV, 
  $M_2=2\times 10^6$ GeV, $M_3=10^6$ GeV, $M_4 = 605$ GeV, 
  $\epsilon_1 \simeq 4.2 \times 10^{-6}$, $\tilde m_1 ^{(3G)}=2.7\times
  10^{-5}$ eV and $\tilde m_1 ^{(4G)}=3.6\times
  10^{-5}$ eV.\\

Although a complete scan of the allowed parameter space varying both the values of the RH neutrino masses and the Yukawa couplings has not been done we mention that it is not completely trivial to satisfy all constraints when we vary the parameters.  For example, for fixed values of the Yukawa couplings, if we let the second lightest RH neutrino be heavier then the heavier LH neutrino becomes lighter than $M_Z/2$. On the other hand, again for fixed values of the Yukawa couplings, when the masses of the two heaviest RH neutrinos increase or decrease, it becomes difficult to have acceptable light LH neutrino masses. Nevertheless, a more detailed anaylsis can provide other spectra for the RH neutrinos which are also acceptable.\\

\begin{figure}[h]
 \begin{center}\begin{tabular}{c}
\includegraphics[width=12cm]{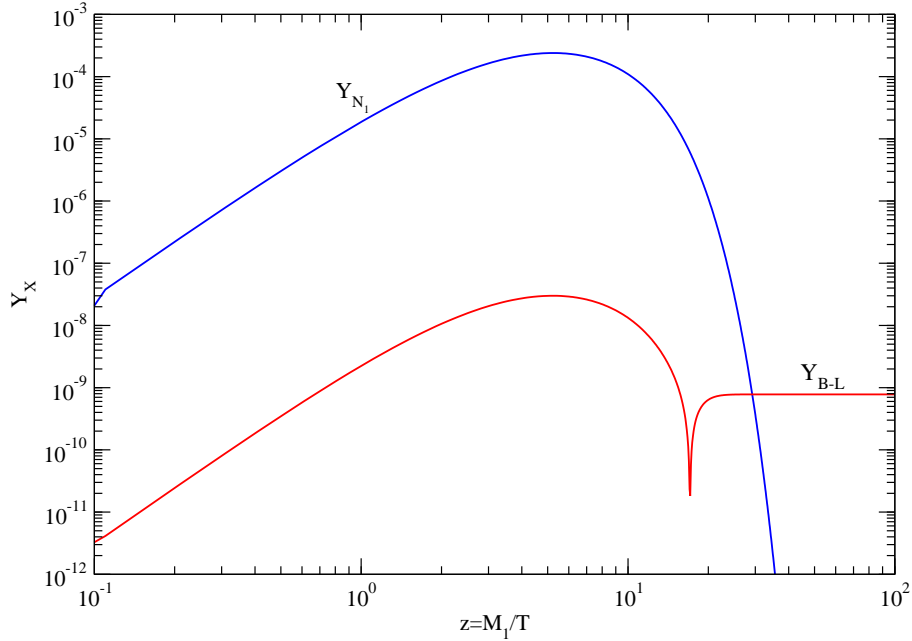}\end{tabular}
\caption{Abundance $Y_{N_1}$ and the baryon asymmetry $Y_{B-L}$ 
for the four generation model.} \label{result4G}
\end{center}
\end{figure}

 Figure  \ref{result4G} represents the solution of the BE, 
 abundance and $B-L$ asymmetry, as a function of $z$. 
 The generated value of the baryon asymmetry is
  $\eta_B \simeq 6\times 10^{-10}$.
  Applying the see-saw mechanism to our model for the chosen values of the 
  parameters, we obtain a heavy left-handed neutrino with a mass above $48$ GeV,
   which is consistent with the fourth generation constraints (see section
   \ref{4gen})
  and the three light neutrino  masses are of the order of $10^{-1}$ eV to a few 
  $10^{-7}$ 
  eV.

\vspace{1cm}
We have presented the solutions to the coupled system of Boltzmann equations
for our TeV scale model of thermal leptogenesis. We have carefully considered 
the
effect of the interactions involving the heavy fourth generation leptonic
 fields
and consistently written the BE which contribute to the final baryon asymmetry 
together
with the appropriate conversion factor from the lepton asymmetry to the
 baryonic one.
Our results show that in this simple extension of the Standard Model it 
is possible to produce the right amount of the BAU in a generic way.


\section*{Acknowledgments}

We would like to thank S. Davidson for extremely useful discussions.
 M. L. thanks the Michigan Center for Theoretical Physics and the
Laboratoire de Physique Th\'eorique, Universit\'e de Paris XI-Orsay 
for hospitality during the completion
of this work. H. A. would like to thank M. Plumacher for useful 
communications.

\end{document}